\documentclass[10pt]{IEEEtran}
\IEEEoverridecommandlockouts
\usepackage{svg}
\usepackage{amsmath,amssymb,amsfonts}
\usepackage{algorithmic}
\usepackage{graphicx}
\usepackage{textcomp}
\usepackage{xcolor}
\def\BibTeX{{\rm B\kern-.05em{\sc i\kern-.025em b}\kern-.08em
    T\kern-.1667em\lower.7ex\hbox{E}\kern-.125emX}}
\usepackage[utf8]{inputenc}
\usepackage{datetime}
\usepackage{upgreek}
\usepackage{dblfloatfix}    
\usepackage{subfigure}
\renewcommand{\figurename}{Fig.}
\usepackage{multirow}
\usepackage{booktabs}
\usepackage{caption}
\usepackage{xcolor,soul,lipsum}
\usepackage{bm}
\usepackage{makecell} 
\usepackage{xr-hyper}
\usepackage{hyperref}
\hypersetup{
    colorlinks=true,
    linkcolor=blue,
    filecolor=magenta,      
    urlcolor=cyan,
}
\usepackage{pifont} 

\usepackage{threeparttable}
\makeatother
\usepackage[hyphenbreaks]{breakurl}
\makeatletter
\usepackage[style=ieee, backend=biber]{biblatex}
\usepackage{upgreek} 

\begin{document}

\title{Unraveling the paradox \\ of intensity-dependent DVS pixel noise\\
\thanks{RG supported by Swiss National Science Foundation grant SCIDVS (200021\_185069). We thank B. Linares-Barranco for helpful comments.}
}

\author{\IEEEauthorblockN{Rui Graca, Tobi Delbruck}
\IEEEauthorblockA{\textit{\\Sensors Group, Inst. of Neuroinformatics, UZH-ETH Zurich, 
Zurich, Switzerland} \\
rpgraca,tobi@ini.uzh.ch, \url{https://sensors.ini.uzh.ch}}
}

\maketitle



\begin{abstract}
Dynamic vision sensor (\textbf{DVS}) event camera output is affected by noise, particularly in dim lighting conditions. 
A theory explaining how photon and electron noise affect DVS output events has so far not been developed. Moreover, there is no clear understanding of how DVS parameters and operating conditions affect noise.
There is an apparent paradox between the real noise data observed from the DVS output and the reported noise measurements of the logarithmic photoreceptor. While measurements of the logarithmic photoreceptor predict that the photoreceptor is approximately a first-order system with RMS noise voltage independent of the photocurrent, DVS output shows higher noise event rates at low light intensity.
This paper unravels this paradox by showing how the DVS photoreceptor is a second-order system, and the assumption that it is first-order is generally not reasonable. As we show, at higher photocurrents, the photoreceptor amplifier dominates the frequency response, causing a drop in RMS noise voltage and noise event rate. We bring light to the noise performance of the DVS photoreceptor by presenting a theoretical explanation supported by both transistor-level simulation results and chip measurements. 

\end{abstract}

\section{Introduction}
\label{sec:abstract}
\begin{figure}[t]
    \centering
    \includegraphics[width=\columnwidth,]{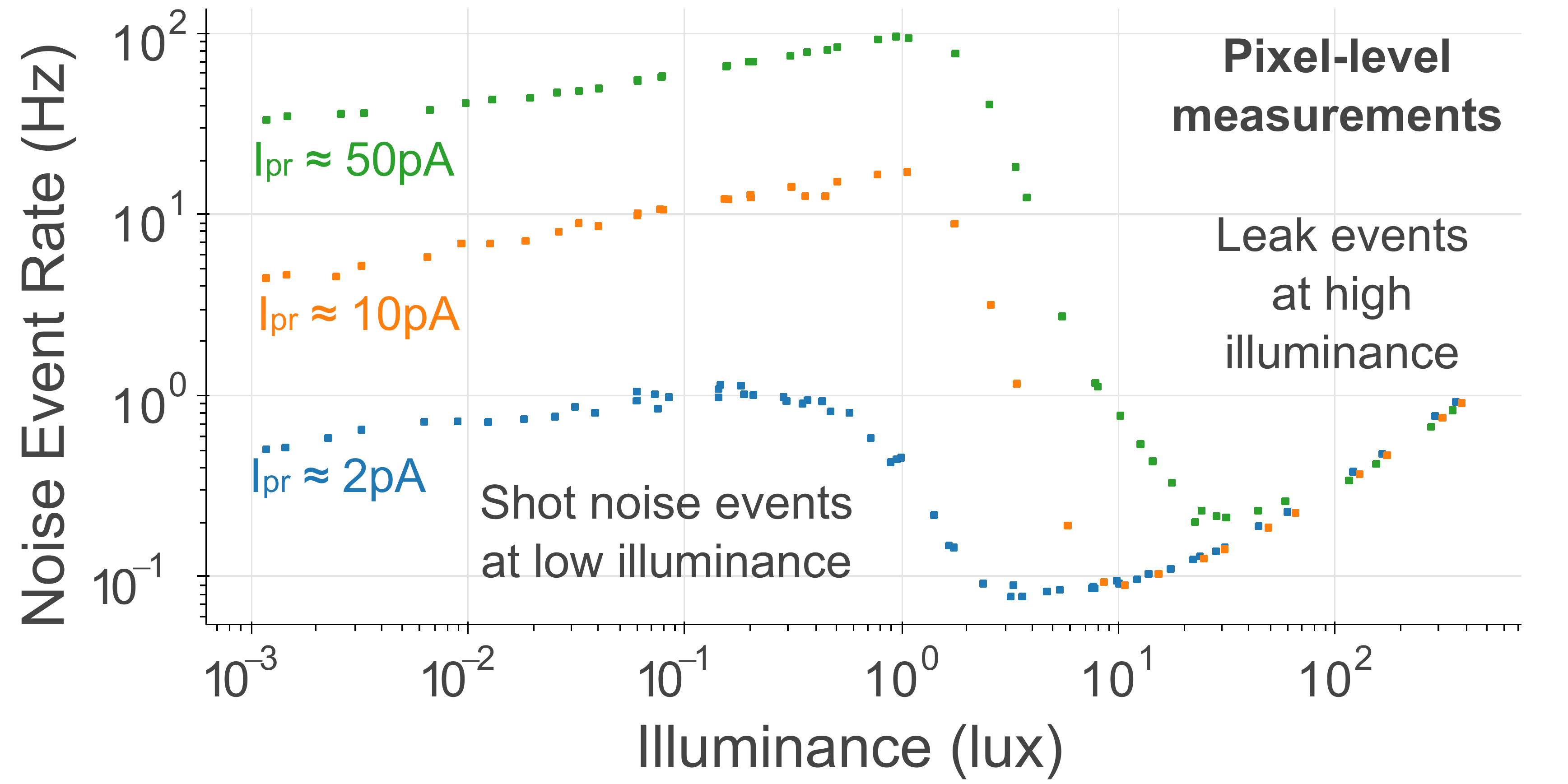}
    \caption{Noise event rate vs. photocurrent for different values of $I_{\text{pr}}$, obtained from DAVIS346 test pixel measurements. The noise event rate is lower at high illuminance, unlike what would be expected from measurements in Fig.~\ref{fig:fig3_log_pr_noise_psd}. This paradox is unravelled in Sec. \ref{sec:unraveling_paradox}.}
    \label{fig:measurements_event_rate}
\end{figure}

\begin{figure*}[t]
    \centering
    \includegraphics[width=\textwidth]{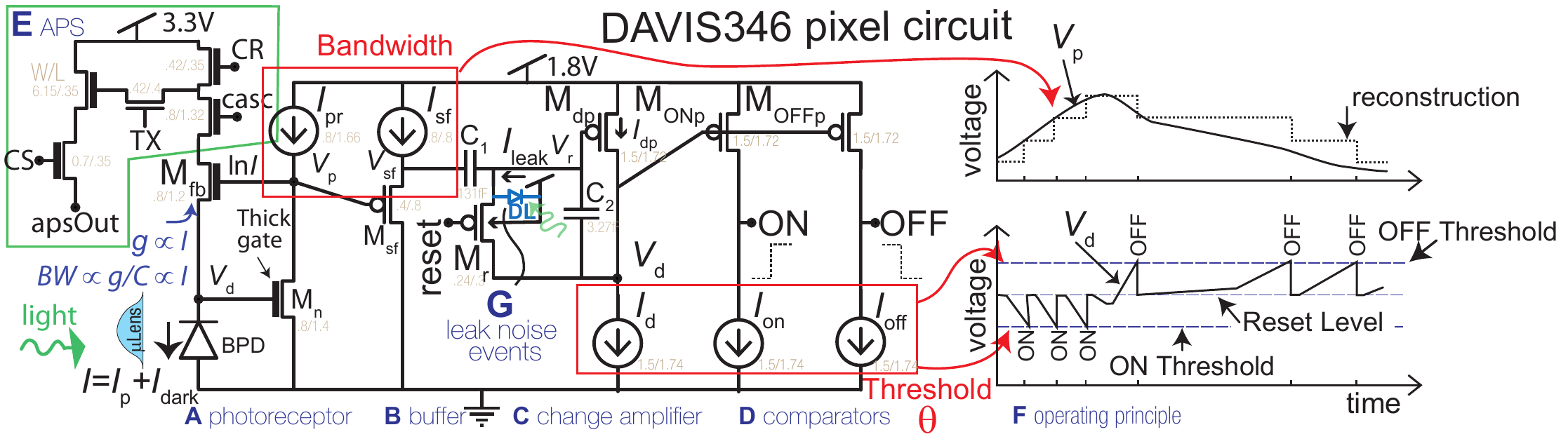}
    \caption{DAVIS346 pixel circuit~\cite{davis346-Taverni2018}. The active logarithmic photoreceptor (\textbf{A}) is buffered by a source-follower (\textbf{B}), which drives a cap-feedback change amplifier (\textbf{C}), which is reset on each event by a low-going \textit{reset} pulse. Comparators (\textbf{D}) detect ON and OFF events as seen in \textbf{F}. Periodic leak events result from junction and parasitic photocurrent $I_\text{leak}$ in diode \textsf{DL} (\textbf{G}). 
    APS output (\textbf{E}) lets us easily measure absolute intensity.}
    \label{fig:fig1-davis-circuit}
\end{figure*}

The DVS pixel (Fig.~\ref{fig:fig1-davis-circuit})~\cite{dvs128,davis346-Taverni2018} is sensitive to the temporal contrast (\textbf{TC}) in light intensity\footnote{TC is the normalized temporal variation in illuminance, or equivalently, the variation in $\ln(I)$.}, producing an ON event when the light intensity increases by a given relative $\theta_\text{ON}$ threshold since the last event, and an OFF event when the light intensity decreases by a given relative $\theta_\text{OFF}$  threshold since the last event (Fig.~\ref{fig:fig1-davis-circuit}F). Ideally, this should result in no events generated when the scene is not changing, and in a deterministic number and timing of events in response to a given change in the scene. However, DVS output includes background activity which is not the result of any changes in the scene~\cite{sony-prophesee-2020}. This background activity is intensity-dependent and abruptly transitions from a low rate of junction-leakage \textit{leak events} under bright lighting~\cite{dvs-temperature-bias} to a high rate of \textit{shot noise events} with dim lighting~\cite{Hu2021-v2e}, as shown in Fig.~\ref{fig:measurements_event_rate} and also reported in~\cite{sony-prophesee-2020}. 

DVS leak events are deterministic and periodic, and they are easy to explain from the pixel reset switch junction leakage~(\textcolor{blue}{\textsf{DL}} in Fig.~\ref{fig:fig1-davis-circuit}G \cite{dvs-temperature-bias}). Shot noise events are caused by random fluctuations in the photoreceptor output voltage. This fluctuation is due to the random arrival and departure times of uncorrelated electrons and photons from circuit nodes~\cite{sarpeshkar1993whitenoise,rose2013vision}. In this paper, we will refer to the shot noise events simply as noise events.

Both leak and noise make signal interpretation more difficult, and they use unnecessary bandwidth in the readout circuitry. The output can be denoised using additional post-processing~\cite{Delbruck2008-frame-free}, but this does not decrease the burden of noise on the DVS arbiter/readout circuitry. Denoising in the pixel array was also proposed in~\cite{Li-vlsi-symp-2019}, at the cost of increasing the pixel size.

Noise performance can be optimized using automatic feedback control of DVS parameters~\cite{delbruck2021feedbackcontrol}, resembling similar strategies of adaptation used in both biological vision and frame-based cameras.

\begin{figure}[h]
    \centering
    \includegraphics[width=\columnwidth]{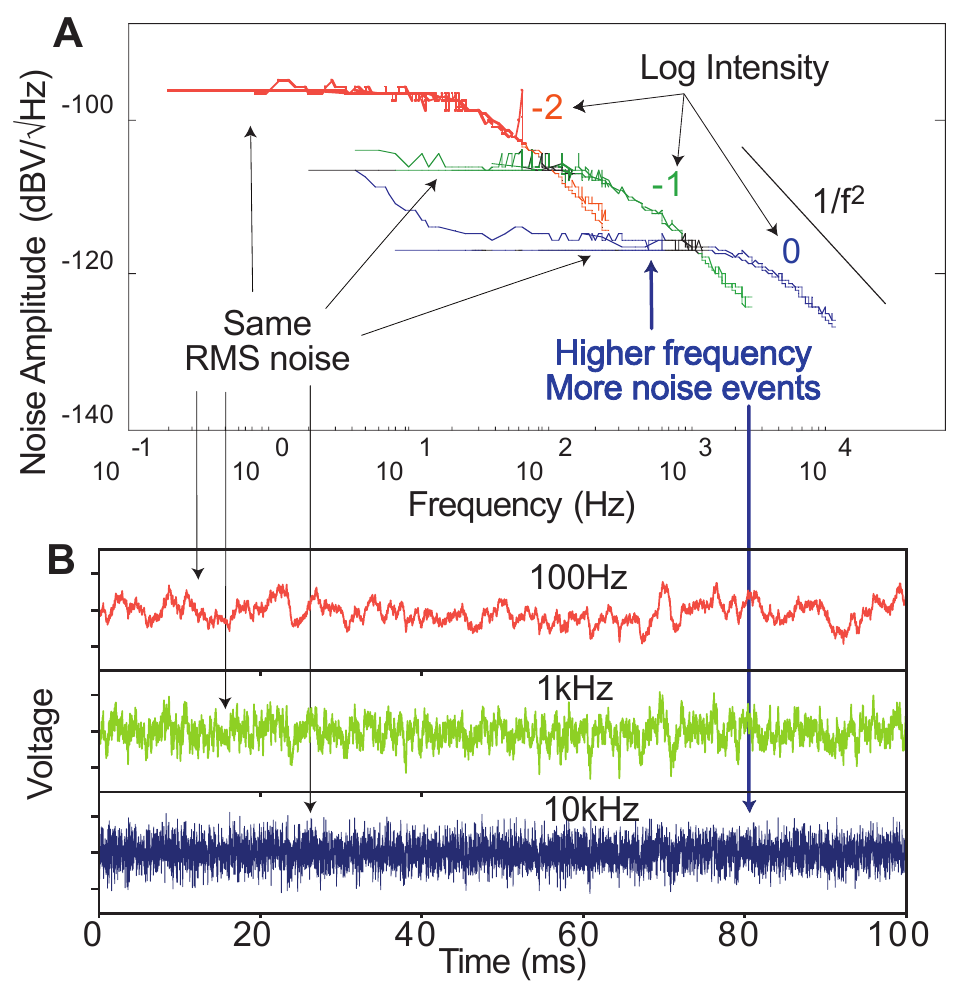}
    \caption{\textbf{A:} Measured PSD of log photoreceptor noise for various intensities (from \cite{Delbruck1994-adaptive-photoreceptor}). \textbf{B:} Simulated white noise with identical total noise power (RMS) but different bandwidth, as in \textbf{A}. }
    \label{fig:fig3_log_pr_noise_psd}
\end{figure}

 Current knowledge about the noise performance of the DVS pixel relies on measurements reported for the logarithmic photoreceptor~\cite{Delbruck1994-adaptive-photoreceptor}, which inspired the current DVS photoreceptor. Fig.~\ref{fig:fig3_log_pr_noise_psd}  shows that it behaves as an intensity-dependent first-order low-pass circuit. In this circuit, the total integrated noise power (i.e., mean square fluctuation) is independent of the photocurrent.

This behavior would lead to the conclusion that the DVS noise event rate should \textit{increase} with photocurrent, since noise in the time domain has the same amplitude, but changes faster with increased intensity (Fig.~\ref{fig:fig3_log_pr_noise_psd}B). This is, however, the \textit{opposite} of what is observed~\cite{sony-prophesee-2020}, as shown in Fig.~\ref{fig:measurements_event_rate}. As we show in Sec.~\ref{sec:measurements}, the DVS photoreceptor is \textbf{second-order}, and not first-order, and for higher light intensities, photoreceptor noise is \textbf{filtered} by the photoreceptor output stage, \textbf{reducing} RMS noise voltage and consequently \textbf{noise event rate}.

Only one DVS pixel~\cite{yang2015dynamicvision} was reported with considerations about noise in the pixel design. However, the authors only report simulation results and no sound theoretical understanding or noise measurements. Moreover, the aim of the authors was to minimize the photoreceptor RMS noise voltage, and the noise event rate does not seem to be considered. There is no clear understanding reported so far about how the photoreceptor RMS noise voltage is related to DVS pixel noise event rate - which should be the ultimate variable to optimize in pixel design.

In this paper, we describe the DVS noise variation with light intensity in Sec.~\ref{sec:measurements}, unravelling the apparent paradox suggested by measurements in~\cite{Delbruck1994-adaptive-photoreceptor}, and also discuss how far from the fundamental photon shot-noise limits the DVS photoreceptor is at several operating conditions. In Sec.~\ref{sec:quantifying} we discuss and report a novel observation of the relation between DVS pixel noise event rate, photoreceptor noise voltage, and event threshold.

\section{Noise variation with light intensity}
\label{sec:measurements}

\subsection{Photoreceptor noise: Measurement and simulation results}
\label{sec:photoreceptor_noise_measurements}

\begin{figure}[t]
    \centering
    \includegraphics[width=\columnwidth]{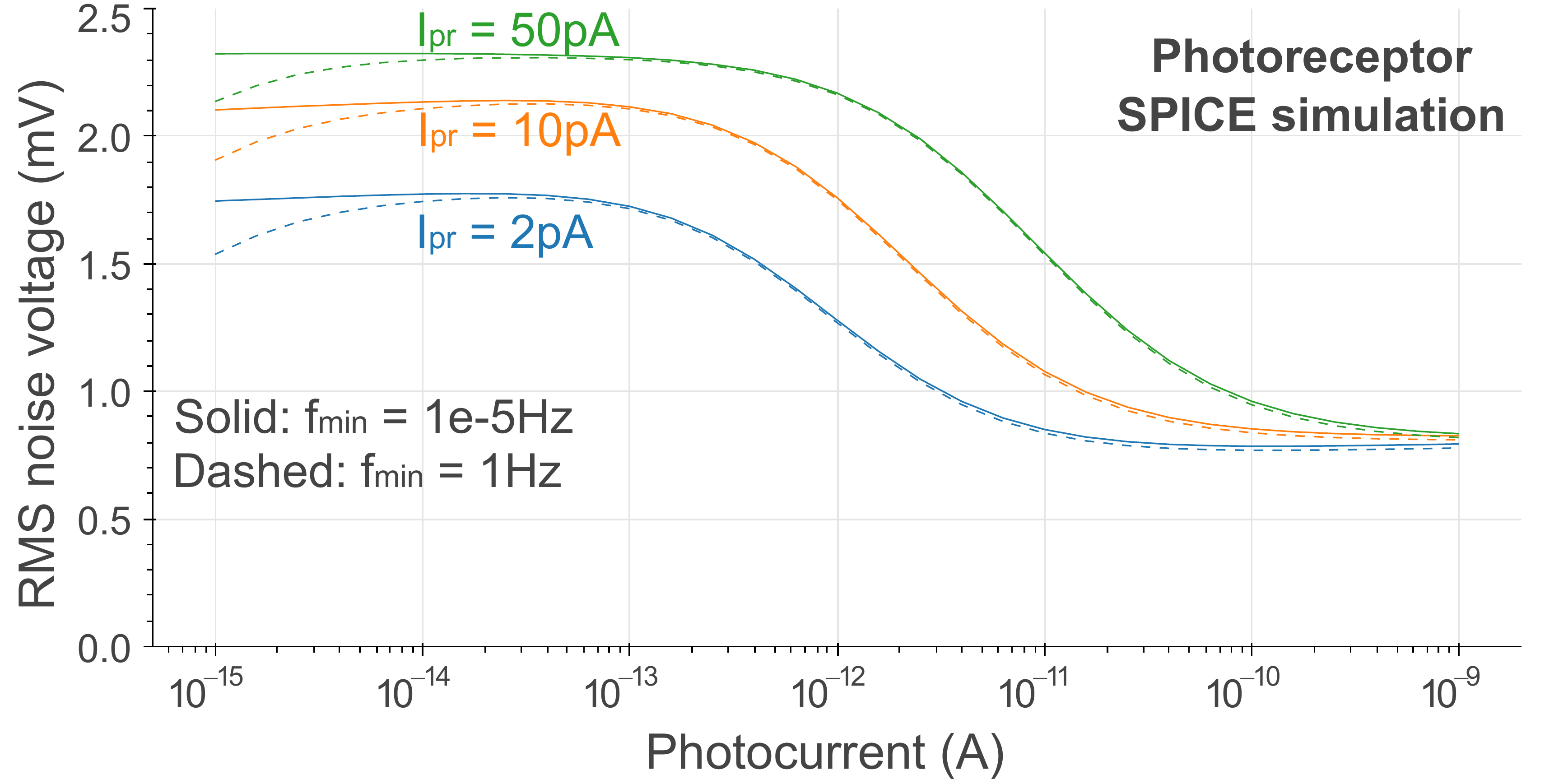}
    \caption{RMS noise voltage at the photoreceptor vs. photocurrent for different values of $I_{\text{pr}}$, obtained from SPICE simulation of a DAVIS346 pixel. Dashed lines are obtained by increasing the minimum noise frequency considered to 1 Hz, the same value used in the chip measurements.}
    \label{fig:fig04_noise_voltage_rms_vs_photocurrent}
\end{figure}

The solid lines in Fig.~\ref{fig:fig04_noise_voltage_rms_vs_photocurrent} shows the photoreceptor RMS noise voltage vs. photocurrent, obtained from SPICE simulations on the photoreceptor of a DAVIS346 chip for three different values of $I_\text{pr}$. Contrary to what is predicted from measurements in the logarithmic photoreceptor~\cite{Delbruck1994-adaptive-photoreceptor}, the output RMS noise voltage is strongly dependent on the photocurrent.

For low photocurrents, the photoreceptor output RMS noise voltage is flat at a level dependent on $I_\text{pr}$, but at a threshold photocurrent, dependent proportionally to $I_\text{pr}$, the photoreceptor output RMS noise voltage drops to a flat level independent of $I_\text{pr}$.

\begin{figure}[t]
    \centering
    \includegraphics[width=\columnwidth]{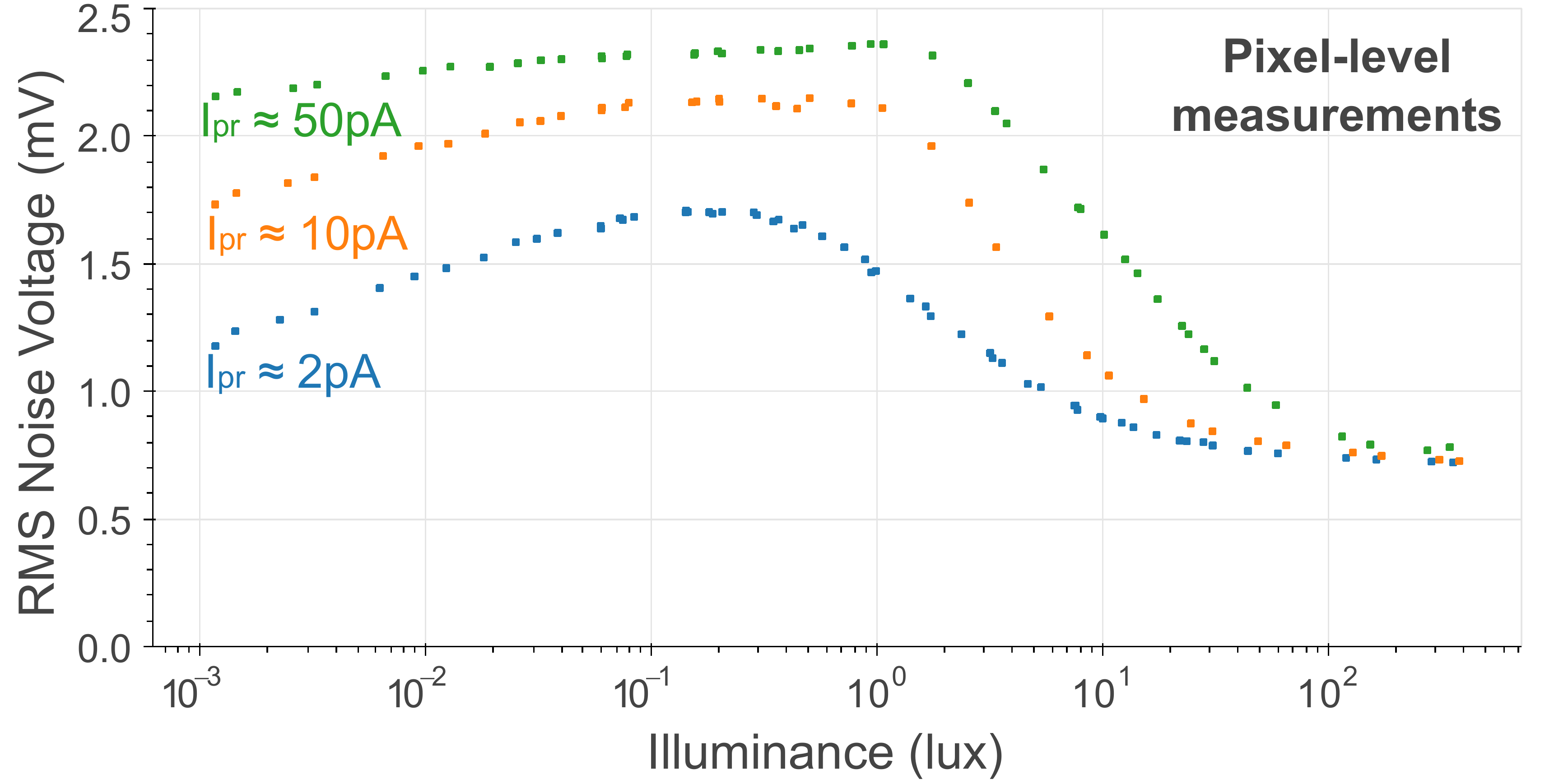}
    \caption{RMS noise voltage at the photoreceptor vs. photocurrent for different values of $I_{\text{pr}}$, obtained from a DAVIS346 test pixel measurement.}
    \label{fig:measurements_vrms}
\end{figure}

\begin{figure}[t]
    \centering
    \includegraphics[width=\columnwidth,]{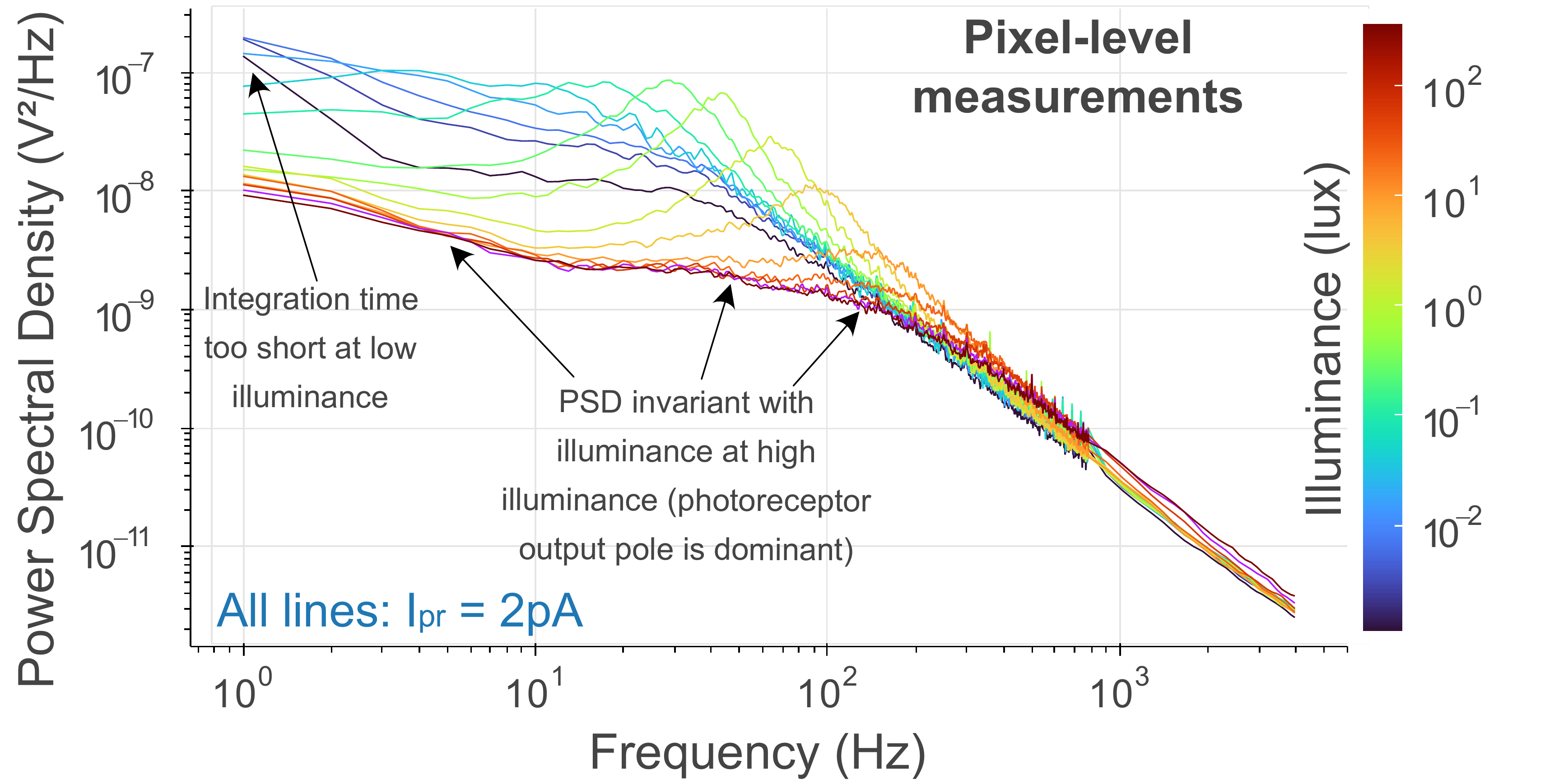}
    \caption{PSD plots obtained from DAVIS346 chip measurements at pixel level for different illuminances for a low setting of $I_{\text{pr}}$.}
    \label{fig:measurements_psd}
\end{figure}

Fig.~\ref{fig:measurements_vrms} shows similar results are obtained from measurements on a DAVIS346 test pixel. The only significant difference is that the chip measurements show a decrease in RMS noise voltage for low illuminance. The PSD spectra obtained from the chip measurements in Fig.~\ref{fig:measurements_psd} show that this difference arises because the integration time used in the measurements is too brief. Increasing the minimum noise frequency used in SPICE simulation results in similar behavior, as shown in the dashed lines in Fig.~\ref{fig:fig04_noise_voltage_rms_vs_photocurrent}.
We come back to interpret these results in Sec.~\ref{sec:unraveling_paradox} after we consider the noise sources.

\subsection{Noise sources}
\label{sec:noise_sources}

To understand the results in Figs.~\ref{fig:fig04_noise_voltage_rms_vs_photocurrent} and \ref{fig:measurements_vrms}, we consider the various noise sources.
Noise at the DVS photoreceptor can be divided into current shot and thermal noise introduced at the input node, $V_\text{d}$, and current noise introduced at the output node, $V_\text{p}$. Noise introduced at the input node consists of shot noise in the photocurrent (due to photon and dark current shot noise) and to the noise current added by $M_\text{fb}$. Noise added at $V_\text{p}$ consists of the noise current in $M_\text{n}$ and the PMOS transistor supplying $I_\text{pr}$. SPICE simulation results show that shot noise dominates in all transistors, and that other noise sources, such as 1/f noise or the noise added by the on-chip bias current generator, are not significant contributors to the overall noise. 

\begin{figure}[t]
    \centering
    \includegraphics[width=\columnwidth]{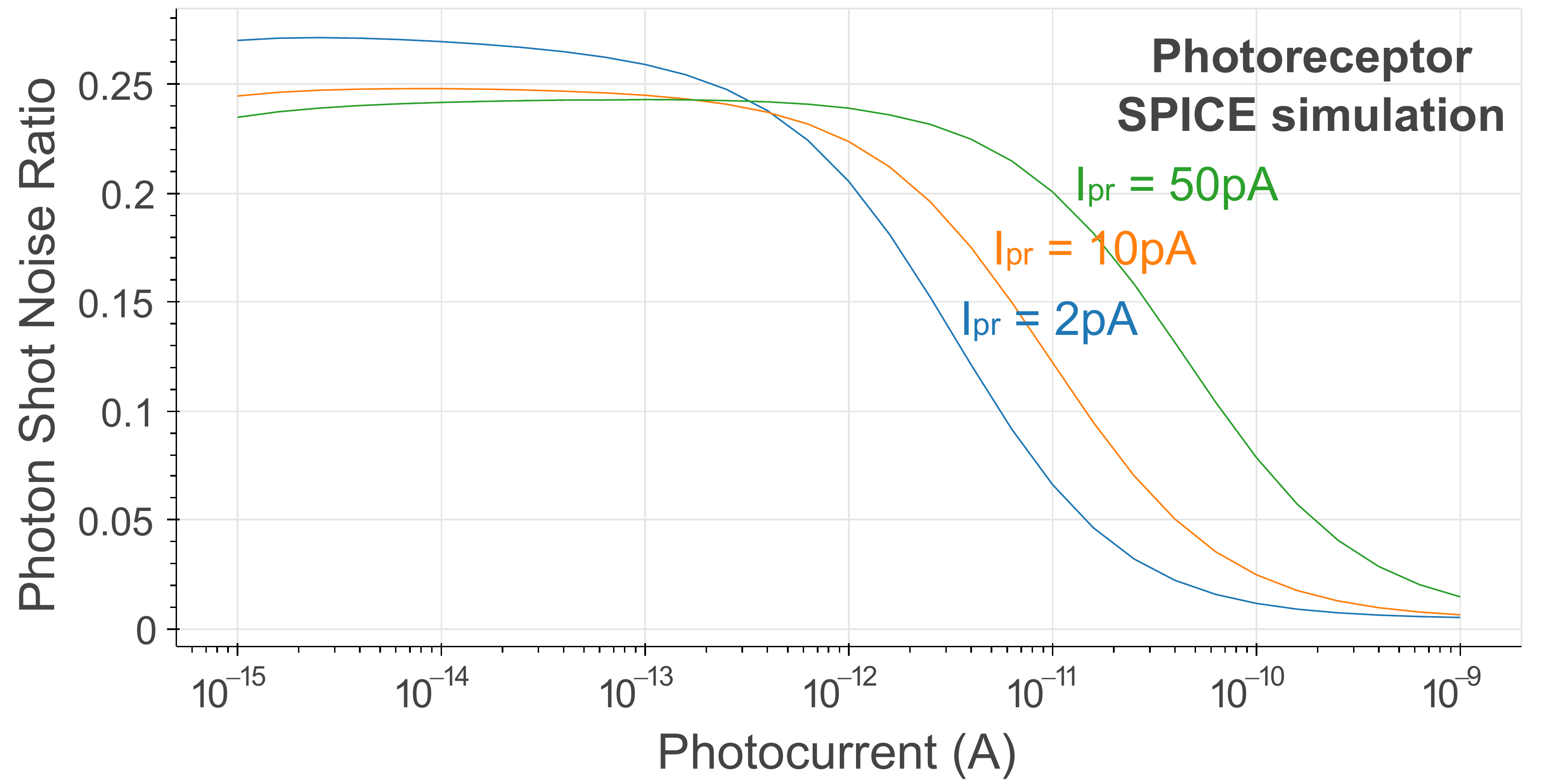}
    \caption{Ratio of the contribution of photon shot noise to the total noise power at the photoreceptor output vs. photocurrent, for different values of $I_{\text{pr}}$, obtained from SPICE simulations.}
    \label{fig:shot_noise_ratio}
\end{figure}

Figure~\ref{fig:shot_noise_ratio} shows the fraction of the output noise power caused by photon shot noise. We see that it is about one quarter of the total noise power for lower photocurrents (with each device introducing about one quarter of the total noise), but it falls to close to zero at high photocurrents. While it is clear that the theoretical limit for noise performance implicates that no noise other than photon shot noise is present, these results alone are not sufficient to determine how close to the ideal noise performance the DVS pixel is operating for two reasons: First, the ideal situation should take into consideration the ideal bandwidth required for a specific scene and application, which is not considered here; and second, as we will show in Sec.~\ref{sec:quantifying}, DVS noise performance is better quantified by the noise event rate. 

\subsection{Noise event rate}
\label{sec:noise_event_rate}

Fig.~\ref{fig:measurements_event_rate} shows the measured noise event rate for the same conditions as in Fig.~\ref{fig:measurements_vrms}. The results are consistent with what was reported at the pixel array level~\cite{sony-prophesee-2020}.  For high photocurrents, the background activity is dominated by intensity-dependent periodic leak events. For photocurrents lower than a bias-dependent threshold, the noise event rate increases very significantly, by an amount dependent on $I_\text{pr}$.

\subsection{Unravelling the paradox}
\label{sec:unraveling_paradox}

The RMS noise measurements obtained in Fig.~\ref{fig:measurements_vrms} clearly show that there is a sudden change in RMS noise voltage at a bias-dependent illuminance threshold. Moreover, above this illuminance threshold, there is a sudden decrease in the noise event rate. 

Observing the PSD curves obtained from the chip measurements in Fig.~\ref{fig:measurements_psd}, we see the behavior of a second-order system, not the first-order behavior observed for the logarithmic photoreceptor in Fig.~\ref{fig:fig3_log_pr_noise_psd}.

For higher illuminance, the pole introduced by the output node starts to dominate, and the photocurrent-dependent noise introduced at the input node is filtered out. This is also the reason the photon noise contribution to the total noise falls to almost zero at high photocurrents as seen in Fig.~\ref{fig:shot_noise_ratio}.

\section{Quantifying performance by noise event rate}
\label{sec:quantifying}

The results in Fig.~\ref{fig:measurements_event_rate} clearly show that for illuminance below about 1\,lux, there is a significant increase in the noise event rate with photocurrent. 
The noise event rate increases more than the total RMS noise increases (Fig.~\ref{fig:measurements_vrms}). The reason for this excess increase can be be easily explained from the results in Fig.~\ref{fig:fig3_log_pr_noise_psd}: Noise with the same total power but higher frequency components will change faster, resulting in higher noise event rates.

The photoreceptor RMS noise has been used as a metric of the DVS noise performance~\cite{yang2015dynamicvision}. It is easy to measure and interpret, and we can easily quantitatively compare it with the TC in log intensity units~\cite{dvs128}: Using typical values for transistor parameters and temperature, an RMS noise voltage of 2\,mV (similar to what is observed at low light intensity) corresponds to a TC of 0.056 log units, and an RMS noise voltage of 0.7\,mV (similar to what is observed at high light intensity) corresponds to a TC of 0.02 log units. 
The event threshold\footnote{We can assume ON and OFF thresholds are the same.} $\theta$ must be some multiple of the TC to prevent high noise rate and large numbers of hot pixels caused by outlier pixels.
The typical $\theta$ used with DVS are between 0.15 and 0.5. $\theta=0.15$ corresponds to 2.7X the TC for dim lighting to to 7.5X the TC for bright lighting. 
However, this does not tell the whole story about noise performance in the DVS pixel. Ultimately, we are interested in the noise event rate at the DVS pixel output. The noise event rate is a function not only of the RMS noise voltage, but also of the distribution of noise power in frequency and amplitude. Moreover, other DVS parameters, such as the event threshold, the refractory period, and the source follower bias current, will influence the noise event rate. The source-follower buffer (Fig~\ref{fig:fig1-davis-circuit}B) and the change amplifier (Fig~\ref{fig:fig1-davis-circuit}C) both filter and introduce noise themselves.

\begin{figure}[t]
    \centering
    \includegraphics[width=\columnwidth]{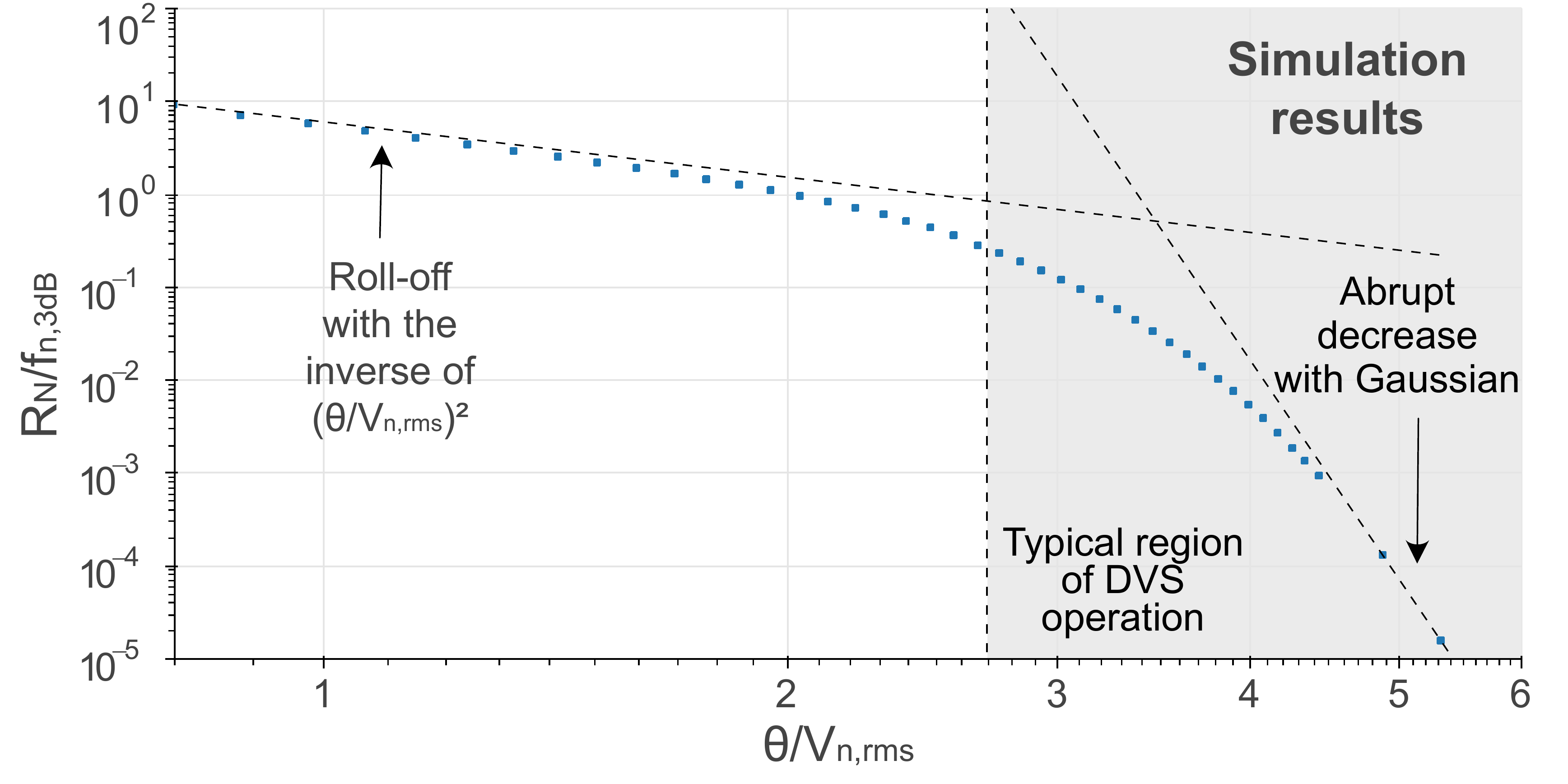}
    \caption{Noise event rate  normalized by photoreceptor cut-off frequency, versus event threshold normalized by photoreceptor RMS noise voltage. Assumes a photoreceptor modeled as first-order low-pass as in Fig.~\ref{fig:fig3_log_pr_noise_psd}.}
    \label{fig:eps_vs_threshold}
\end{figure}

We developed a simplified pixel model to understand how RMS noise and frequency distribution result in a noise event rate. We assume the DVS pixel is a first-order low-pass system (as seen above, this is approximately valid for low photocurrents), and that the refractory period is zero. Fig.~\ref{fig:eps_vs_threshold} shows the results of a numerical simulation using this model. The model predicts that for a given ratio between the event threshold $\theta$ and RMS noise $V_\text{n,rms}$, the resulting noise event rate $R_n$ is proportional to the PSD cut-off frequency $f_\text{n,3dB}$. It also predicts two regions with different behaviors, depending on the value of $r=\theta/V_\text{n,rms}$. For $r<2$, the noise event rate rolls off as $r^2$. This behavior occurs because as $\theta$ is further decreased, more spectral components at frequencies higher than $f_\text{n,3dB}$ get enough power to trigger events. For $\theta$ larger than $2V_\text{n,rms}$ (representing typical usage), $R_n$ decreases very steeply: It falls off with the exponential tail of the Gaussian distribution of amplitudes, since the generation of a noise event depends on the noise voltage crossing a specific threshold. 

This observation explains the observed steep increase of noise event rate as $\theta$ is decreased and suggest that a possible sweet spot for threshold lies at some multiple of $r=1$, suggesting it is possible to develop a procedure for principled selection of the optimum $\theta$ given measurement of photocurrent (e.g. via DAVIS APS reading) and control of $I_\text{pr}$ bias current.

\section{Conclusion}
\label{sec:conclusion}
This paper explains how DVS photoreceptor noise depends on photocurrent and biasing. We present evidence based on simulation results and chip measurements at the pixel level that is consistent with what would be theoretically predicted from circuit analysis at the pixel level. Our results indicate the behavior of a second-order photoreceptor transfer function. Our results explain the steep increase in noise event rate at low light intensities~\cite{sony-prophesee-2020} and the behavior of noise event rate with event threshold. 




\renewcommand*{\bibfont}{\footnotesize}
\printbibliography




\end{document}